\documentclass[aps,prb,twocolumn,superscriptaddress]{revtex4}

\usepackage{amsmath}
\usepackage{graphicx}
\usepackage{dcolumn}
\usepackage{bm}

\begin{document}

\title{Rashba spin splitting of $L$-gap surface states on Ag(111) and Cu(111)}

\author{Koichiro Yaji}
\email{yaji@issp.u-tokyo.ac.jp}
\affiliation{Institute for Solid State Physics, The University of Tokyo, 5-1-5 Kashiwanoha, Kashiwa, Chiba 277-8581, Japan}

\author{Ayumi Harasawa}
\affiliation{Institute for Solid State Physics, The University of Tokyo, 5-1-5 Kashiwanoha, Kashiwa, Chiba 277-8581, Japan}

\author{Kenta Kuroda}
\affiliation{Institute for Solid State Physics, The University of Tokyo, 5-1-5 Kashiwanoha, Kashiwa, Chiba 277-8581, Japan}

\author{Ronghan Li}
\affiliation{Shenyang National Laboratory for Materials Science, Institute of Metal Research, Chinese Academy of Science, School of Materials Science and Engineering, University of Science and Technology of China, 110016, Shenyang, China}
\affiliation{Department of Condensed Matter Physics, Weizmann Institute of Science, 7610001 Rehovot, Israel}

\author{Binghai Yan }
\affiliation{Department of Condensed Matter Physics, Weizmann Institute of Science, 7610001 Rehovot, Israel}

\author{Fumio Komori}
\email{komori@issp.u-tokyo.ac.jp}
\affiliation{Institute for Solid State Physics, The University of Tokyo, 5-1-5 Kashiwanoha, Kashiwa, Chiba 277-8581, Japan}

\author{Shik Shin}
\affiliation{Institute for Solid State Physics, The University of Tokyo, 5-1-5 Kashiwanoha, Kashiwa, Chiba 277-8581, Japan}

\begin{abstract}

Spin-resolved band structures of {\it L}-gap surface states on Ag(111) and Cu(111) are investigated by spin- and angle-resolved photoelectron spectroscopy (SARPES) with a vacuum-ultra-violet laser. 
The observed spin textures of the Ag(111) and Cu(111) surface states agree with that expected by the conventional Rashba effect. The Rashba parameter of the Ag(111) surface state is estimated quantitatively and is 80$\%$ of that of Cu(111). 
The surface-state wave function is found to be predominantly of even mirror-symmetry with negligible odd contribution by SARPES using a linearly polarized light. 
The results are consistent with our theoretical calculations for the orbital-resolved surface state.

\end{abstract}


\maketitle

Spin-polarized metallic systems including non-magnetic elements have attracted much attention for decades because novel spin transports and their applications to information technology can be expected by using these systems. 
The spin splitting of a two-dimensional electron system due to the Rashba effect \cite{Rashba_1960, Bychkov_1984} at crystal surfaces, where a potential gradient is naturally built, has been studied intensively as one of the controllable spin-polarized systems. 
The $L$-gap surface states of the noble metal (111) surfaces behave as prototypical two-dimensional free electron gas (2DEG) systems, as shown by Shockley \cite{Shockley_PRB_1939}, with a Rashba-type spin splitting. 
The energy dispersion of 2DEG at the surface in an electric field $E_{z}$ perpendicular to the surface can be expressed as \cite{Rashba_1960, Bychkov_1984} 
\begin{math}
E(k)=\frac{\hbar^2}{2m^{*}}k_{\rm ||}^2 \pm \alpha_{\rm R}k_{\rm ||}
\end{math}
with a term expressing the momentum splitting $\alpha_{\rm R}k_{\rm ||}$. Here, the parameter, $\alpha_{\rm R}$  =  $\frac{\hbar^2E_{z}}{4m^{*2}c^{2}}$, is the Rashba parameter expressing the strength of the Rashba effect, $m^{*}$ the effective mass, $c$ the light velocity. 
This equation represents that a parabolic band of 2DEG is split into two along the in-plane momentum directions. 

The surface Rashba effect was first reported on the clean Au(111) surface \cite{LaShell_PRL_1996}. The importance of the atomic spin-orbit coupling strength was pointed out in the literature. 
Subsequently the spin splitting was confirmed by spin- and angle-resolved photoemission spectroscopy (SARPES) \cite{Hoesch_PRB_2004}, where the energy dispersion of the band is consistent with the original Rashba effect although the spin-polarization directions of two split branches are opposite to those predicted by the original Rashba Hamiltonian. 
The observed value of $\alpha_{R}$ is 0.33 eV\AA\, which is 5 orders of magnitude larger than that expected for 2DEG at the Au(111) surface by considering the surface potential gradient \cite{LaShell_PRL_1996}, while a full-potential first-principles calculation \cite{Henk_PRB_2003,Nicolay_PRB_2001, Reinert_JPCM_2003} later confirmed this value. 
Thus, the origin of the spin splitting in the surface states has not yet fully understood even in simple systems and several possible mechanisms have been suggested: a model of in-plane inversion asymmetry and in-plane potential gradient \cite{Ast_PRL_2007, Premper_PRB_2007, Ast_PRB_2008, Gierz_PRL_2009, Moreschini_PRB_2009}, surface-perpendicular asymmetry of charge density distribution in close proximity to the nuclei \cite{Bihlmayer_SurSci_2006, Nagano_JPCM_2009, Hatta_PRB_2009, Yaji_NatComm_2010, Bentmann_PRB_2011, Hortamani_PRB_2012}, orbital angular momentum of surface-state wave functions \cite{Kim_PRB_2012}, the importance of $d$-orbitals in the surface-state wave functions \cite{Lee_PRB_2012}, relativistic modification of the surface-state wave function \cite{Krasovskii_PRB_2014} and the spin-orbit coupling as perturbation to scalar-relativistic wave functions \cite{Ishida_PRB_2014}. 
Moreover, recent theoretical study proposes these $L$-gap surface states of noble metals (111) are spin-split topological states \cite{Yan_NatCommun_2016}. 

Experimentally, splitting of the {\it L}-gap surface state of the Cu(111) surface was studied by high-resolution angle-resolved photoemission spectroscopy (ARPES) using a vacuum-ultra-violet laser \cite{Tamai_PRB_2013}.  
The splitting is unexpectedly large and just 1/4 of the surface state of Au(111) while the atomic number is 1/3. 
Here, the atomic spin-orbit coupling of Au 6$p$ is more than an order of magnitude lager than that of Cu 4$p$ \cite{SOC}. 
In theories \cite{Reinert_JPCM_2003, Ishida_PRB_2014}, the calculated spin splitting of the Ag(111) surface state is smaller than that of Cu(111). 
These suggest a simple consideration based on the strength of the atomic spin-orbit interaction is not applicable for these systems.

\begin{figure*}
\includegraphics[keepaspectratio=true,scale=1.0]{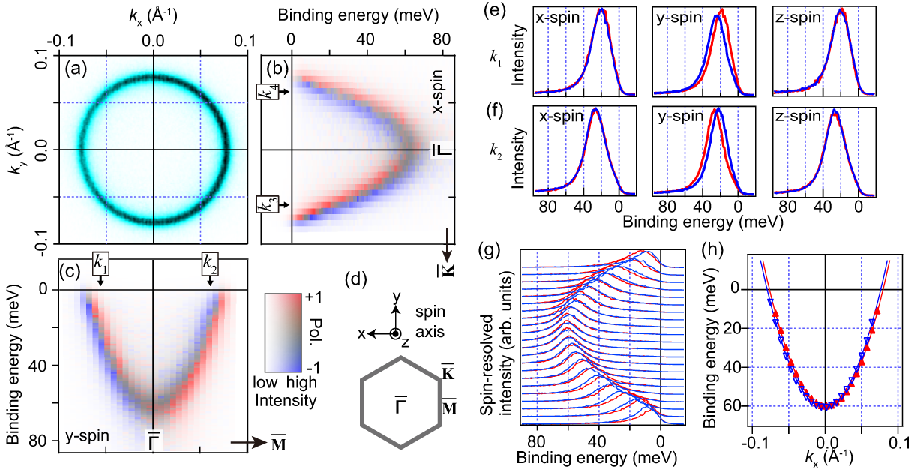}
\caption{\label{fig:epsart1}
(Color online) 
 (a) Constant energy laser-ARPES intensity map at the Fermi level ($E_{\rm F}$) of the Ag(111) surface state recorded with the $p$-polarized light. 
 (b,c) Spin polarization and photoelectron intensity images along the $\bar{\Gamma}\bar{\rm K}$ (b) and $\bar{\Gamma}\bar{\rm M}$ (c) directions in the surface Brillouin zone of the Ag(111) surface measured by laser-SARPES. 
 A color scale gives both amplitude of the spin polarization and the photoelectron intensity. 
(d) Surface Brillouin zone of the face-centered-cubic (111) surface and definition of the spin-polarization direction. 
(e,f) Three-dimensional spin-resolved EDCs at $k_{\rm 1}$ and $k_{\rm 2}$ shown in (c). 
Red and blue curves represent the positive (negative) spin polarization direction depicted in (d). 
(g) Emission angle dependence of the spin-resolved EDCs of the Ag(111) surface state along $\bar{\Gamma}\bar{\rm M}$, where the spin detector is arranged to detect the $y$ component of the spin polarization. 
Red and blue curves represent the positive (negative) spin polarization defined in (d). 
(h) Red filled and blue opened triangles represent peak positions obtained from the spin-resolved EDCs in (g). 
Red and blue curves represent fitting results with free-electron-like parabolas.
}
\end{figure*}

In a conventional model of the Rashba spin splitting, the spin orientation of an electron is locked to the momentum. 
In contrast to this, spin-orbital entanglement in spin-orbit-coupled surface states has been recently revealed; it describes a spin texture coupled to orbital symmetries \cite{Cao_NatPhys_2013, Zhang_PRL_2013}, and was experimentally found in several spin-orbit coupled materials \cite{Yazyev_PRL_2010, Zhu_PRL_2013, Xie_NatComm_2014, Zhu_PRL_2014, Bawden_SciAdv_2015, Maas_NatComm_2016, Ryoo_PRB_2016, Miyamoto_PRB_2016, Kuroda_PRB_2016, Noguchi_PRB_2017, Yaji_NatCommun_2017}. 
Besides, as a consequence of the spin-orbit coupling and mirror symmetry, the spins pointing to the mutually opposite directions are independently locked to the even-odd parity symmetries \cite{Henk_PRB_2003, Yaji_NatCommun_2017} in the mirror plane. 
This indicates that the spin polarization observed by SARPES must be reversed upon switching the light polarization from $p$ to $s$ within dipole transition approximation. 
In the case of the Au(111) surface state, however, absence of the spin reversal upon switching of the linear polarization of the excitation light was previously reported \cite{Jozwiak_NatPhys_2013}. 
Subsequently, Ryoo {\it et al.} have theoretically pointed out that the non-reversal spin polarization of the Au(111) surface state can be caused by the imperfect polarization and non-normal incidence of the light \cite{Ryoo_PRB_2016}. 

In this Rapid Communication, we report first experimental demonstration of the Rashba-type spin splitting of the Ag(111) surface state investigated by SARPES using a vacuum-ultra-violet laser (laser-SARPES). 
We also show the spin splitting of the Cu(111) surface state by laser-SARPES. 
The Rashba parameters of the Ag(111) and Cu(111) surface states are experimentally determined from these results. 
Furthermore, we elucidate that the wave function of the Cu(111) surface state predominantly consists of the symmetric orbital components.

Clean Ag(111) and Cu(111) surfaces were {\it in situ} prepared by repeated cycles of 0.5 keV Ar$^+$ bombardment and subsequent annealing up to 770 K for Ag(111) and 850 K for Cu(111). The surface lattice order was checked by the sharpness of a low energy electron diffraction pattern as a first step. 
Eventually, we judged the effect of any electron scatterings due to the surface impurities and defects from spectral widths of momentum distribution curves observed by ARPES. 
In the ARPES and SARPES measurements, photoelectrons were excited by a quasi-continuous-wave laser with the photon energy of 6.994 eV and were analyzed with a combination of a hemispherical electron energy analyzer (ScientaOmicron DA30L) and twin very-low-energy-electron-diffraction type spin detectors orthogonally placed\cite{Yaji_RSI_2016}. Degree of linear polarization of the light was 97\%. 
The light incident plane is parallel to the (1$\bar{1}$0) plane of the sample that corresponds to the $\bar{\Gamma}\bar{\rm M}$ mirror plane. 
The laser-ARPES and laser-SARPES spectra were taken with instrument energy resolutions of 5 meV and 6 meV, respectively. 
The sample temperature was kept at 12 K during the measurements. 
Calculations of the surface states were carried out by density-functional methods within the Perdew-Burke-Ernzerhof-type generalized gradient approximation \cite{GGA}. A slab model with thirty-seven atomic layers was adopted to simulate the surfaces. 

Figures 1(a)--1(c) display the Fermi surface mapping of the Ag(111) surface state by laser-ARPES and the spin-resolved band mapping along $\bar{\Gamma}\bar{\rm K}$ and $\bar{\Gamma}\bar{\rm M}$ by laser-SARPES. 
The circular shape of the Fermi surface, which is in good agreement with the published literature \cite{Reinert_PRB_2001}, indicates that the Ag(111) surface state exhibits ideal 2DEG nature. 
In Figs. 1(b) and 1(c), we observed the spin polarization perpendicular to both the electron momentum and surface normal; the spin signal in the $x$ ($y$) direction was observed for the band along  $\bar{\Gamma}\bar{\rm K}$ ($\bar{\Gamma}\bar{\rm M}$). 
Here, the $x$ ($y$) direction is parallel to the [11$\bar{2}$] ([$\bar{1}$10]) axis and the $z$ direction corresponds to the surface normal. 
The spin-resolved band images clearly demonstrate the splitting of the Ag(111) surface state, and the spin-polarized branches are oppositely spin-polarized to each other with respect to the $\bar{\Gamma}$ point. 

Figures 1(e) and 1(f) show spin-resolved energy distribution curves (EDCs) at selected {\it k} cuts \cite{Supple1}.
In the $\bar{\Gamma}\bar{\rm M}$ ($\bar{\Gamma}\bar{\rm K}$) direction, the $y$ ($x$) spin polarization is found while there is no spin signal in the $x$ and $z$ ($y$ and $z$) directions. 
These results suggest that the {\it L}-gap surface state of Ag(111) exhibits the tangential spin texture with respect to the circular-shape Fermi surface, which is expected from the conventional Rashba effect.

\begin{figure}
\includegraphics[keepaspectratio=true,scale=0.9]{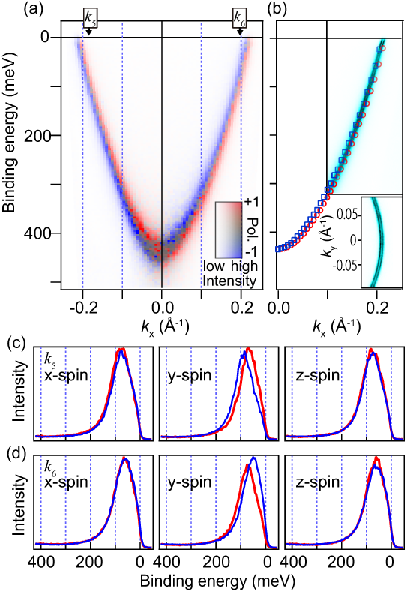}
\caption{\label{fig:epsart2}
(Color online)  (a) Spin polarization and photoelectron intensity images of the Cu(111) surface state along $\bar{\Gamma}\bar{\rm M}$ measured by laser-SARPES. The spin detector is arranged to be sensitive to the $y$ component of the spin polarization [see Fig. 1(d) for the definition of the spin polarization direction]. 
(b) Peak positions of the spin-resolved EDCs together with the laser-ARPES intensity image. Red and blue symbols correspond to mutually opposite spin directions in the $y$ direction. 
In the inset, the constant energy ARPES intensity map at $E_{\rm F}$ is displayed. 
The constant energy image is obtained by the summation of the photoelectron intensity within a 2-meV energy window centered at $E_{\rm F}$. 
(c,d) Three-dimensional spin-resolved EDCs at $k_{\rm 5}$ and $k_{\rm 6}$ shown in (a). Red and blue curves represent the positive (negative) spin polarization direction defined in Fig. 1(d). 
}
\end{figure}

The spin splitting of the Ag(111) surface state is quantitatively analyzed with the spin-resolved EDCs shown in Fig. 1(g). 
We found that the spin-up and spin-down peaks are clearly identified and are inverted with respect to the $\bar{\Gamma} $ point. 
In Fig. 1(h), we plot peak positions of the spin-resolved EDCs. 
The peak position values were fitted by free-electron-like parabolas with three fitting parameters, that is, the Rashba parameter $\alpha_R$, an effective mass and energy of the band bottom. 
From the fitting, we obtain $\alpha_R$ of $\sim$0.031 eV\AA, the effective mass of $\sim$0.38$m_{\rm e}$, and the band bottom energy of $\sim$61 meV, where $m_{\rm e}$ is the free electron mass.
The size of the energy splitting at the Fermi wave vector between the spin-up and spin-down states is $\sim$5 meV. 
The splitting in the momentum direction is estimated to be $\sim$0.003 \AA$^{-1}$. 
These experimental values characterizing the Rashba effect are two times larger than the theoretical prediction by Reinert {\it et al.} \cite{Reinert_JPCM_2003}. and are smaller than that by Ishida \cite{Ishida_PRB_2014}. 

Next, we show the spin splitting of the Cu(111) surface state. 
Figures 2(a) and 2(b) displays spin-resolved and spin-integrated band mappings along $\bar{\Gamma}\bar{\rm M}$ by laser-(S)ARPES. 
In this case, we observed the splitting of the band and the Fermi surface by laser-ARPES, and demonstrated the spin-dependent contrast in the band image by laser-SARPES. 
Using the three-dimensional SARPES measurements, we find the spin polarization only in the $y$ direction (Fig. 2(c,d)). 
Therefore, the spin texture of the Cu(111) surface state can be described by the conventional Rashba effect model. 
We estimated the Rashba parameter to be $\sim$0.038 eV\AA. 
Besides, we found that the size of the energy splitting at the Fermi wave vector is $\sim$16 meV and the energy of the band bottom is $\sim$434 meV, which are in excellent agreement with the former report \cite{Tamai_PRB_2013}. 

Recently, Dil {\it et al.} reported the spin interference in the photoemission process between the spin-lifted branches in the systems of Cu(111) and Sb/Ag(111), where the spin interference rotates the spin vector of the photoelectron such that the measured spin polarization of the photoelectron is not the same as the spin polarization of the Bloch state \cite{Dil_JESRP_2015, Meier_JPCM_2011}. 
Such spin interference is claimed to be possible when the spin-lifted branches are very close to each other and are partly overlapped due to the momentum broadening induced by elastic scattering, associated with point defects in imperfect surfaces for example. 
In contrast, this is not the case in the sufficiently-cleaned surfaces\cite{Fanciulli_PRL_2017}, where we have demonstrated the conventional Rashba-type spin splitting for Ag(111) and Cu(111) as shown above. 

\begin{table} 
\caption{\label{tab:table1} Rashba parameters $\alpha_{\rm R}$ and spin-splitting energies $\mathit{\Delta}\epsilon$ at the Fermi wave vectors of the spin-polarized surface states of noble metal (111)\cite{Rashba_para_calc}.}
  \begin{tabular}{c|c|c|c} \hline
      &$\alpha_{\rm R}$(eV\AA)&$\mathit{\Delta}\epsilon$(meV)&Ref. \\ \hline
      Cu(111)&0.038&16&This work\\ 
      Cu(111) calculation&0.059&26&[\cite{Ishida_PRB_2014}]\\ 
      Ag(111)& 0.031&5&This work\\
      Ag(111) calculation& 0.012&1.9&[\cite{Nicolay_PRB_2001,Reinert_JPCM_2003}]\\
      Au(111)&0.33&110&[\cite{LaShell_PRL_1996}] \\  \hline
   \end{tabular}
\end{table}

Table I gives a comparison of the Rashba parameters of the surface states of Cu(111), Ag(111) and Au(111).  
We have experimentally revealed that the Rashba parameter of Ag(111) is 80$\%$ of that of Cu(111), being consistent with the theoretical predictions \cite{Nicolay_PRB_2001, Reinert_JPCM_2003, Ishida_PRB_2014}. 
According to the recent theoretical studies by Ishida \cite{Ishida_PRB_2014}, the {\it d}$_{\rm z^{2}}$ and {\it d}$_{\rm xz}$ orbital components mixed in the surface-state wave function, in which the {\it p}$_{\rm z}$ orbital component is dominant, play important role in determining the size of the spin splitting. 
In fact, the {\it d}$_{\rm z^{2}}$ and {\it d}$_{\rm xz}$ orbital components in the surface-state of Cu(111) are substantially larger than the corresponding ones of Ag(111) \cite{Ishida_PRB_2014}. 
This scenario is convincible to explain the larger spin splitting of Cu(111) than Ag(111), experimentally found in the present study. It is also possible to attribute the spin splitting of these systems to the topological origin \cite{Yan_NatCommun_2016}. 
The scenario using the topology explains the origin of the Shockley-type surface states, as implied in the early paper by Shockley \cite{Shockley_PRB_1939}. 
However, the topological scenario itself is not helpful to determine the details of the spin-polarized band, such as the spin direction, the spin polarization and the size of the spin splitting quantitatively. 

\begin{figure}
\includegraphics[keepaspectratio=true,scale=1.0]{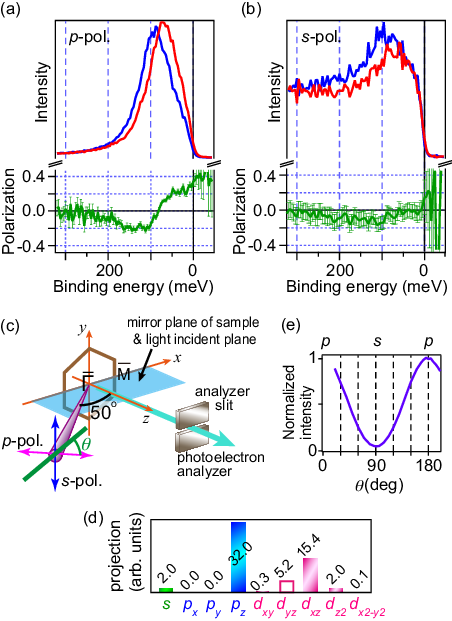}
\caption{\label{fig:epsart3}
(Color online) 
(a,b) (upper panel) Spin-resolved EDCs and (lower panel) the spin polarizations for the $y$-spin component at the wave vectors of $k_{\rm 5}$ shown in Fig. 2(a) measured with (a) $p$- and (b) $s$-polarizations. 
(c) Schematic drawing of the experimental geometry and the definition of the angle $\theta$ as an angle between the electric-field vector of the incident light and the light incident plane. The {\it p}-polarization ({\it s}-polarization) corresponds to $\theta = 0^{\circ}$ and 180$^{\circ}$ ($\theta = 90^{\circ}$). 
The light incident plane is parallel to the mirror plane of the crystal. 
The angle between the light and the analyzer was fixed to 50$^{\circ}$. 
(d) Orbital projection in the mirror plane for the Cu(111) surface state at $E_{\rm F}$. 
Filled (opened) bars represent the even (odd) orbitals. The total weight of the even (odd) orbital contribution is 51.5 (5.5). 
(e) Photoelectron intensity at $k_{\rm 6}$ shown in Fig. 2(a) as a function of $\theta$. 
}
\end{figure}

We examine the spin-orbital entanglement of the Cu(111) surface state. 
The spin-orbital entanglement is elucidated by employing the dipole selection rule of a linearly-polarized light in photoemission. 
Figures 3(a) and 3(b) show spin-resolved EDCs and the spin polarizations measured with $p$- and $s$-polarized lights for the Cu(111) surface state at $k_{\rm 5}$ shown in Fig. 2(a). 
The experimental geometry is displayed in Fig. 3(c). 
We find that the spin polarization is not inverted and the value of the spin polarization is reduced upon switching the linear polarization from $p$ to $s$ in contrast to the previous results in Bi$_2$Se$_3$, Bi(111), W(110) and Bi/Ag(111) \cite{Zhu_PRL_2013, Xie_NatComm_2014, Zhu_PRL_2014, Bawden_SciAdv_2015, Maas_NatComm_2016, Ryoo_PRB_2016, Miyamoto_PRB_2016, Kuroda_PRB_2016, Noguchi_PRB_2017, Yaji_NatCommun_2017}. 

To understand the non-reversal of the spin polarization, we have calculated the projection of the surface-state wave function in the $\bar{\Gamma}\bar{\rm M}$ mirror plane at $E_{\rm F}$ [Fig. 3(d)] and have investigated the light polarization dependence of the photoelectron intensity [Fig. 3(e)]. 
In the mirror plane, we find the symmetric orbitals $p_{\rm z}$ and $d_{\rm xz}$ are dominant while the anti-symmetric part $d_{\rm yz}$ contributes a little.
Whereas the contributions from $p_{\rm x}$, $p_{\rm y}$, $d_{\rm xy}$ and $d_{\rm x^{2}-y^{2} }$ states are negligible, and there are small contributions from  $s$ and $d_{\rm z^2}$. 
Experimentally, the photoelectron intensity with the $s$-polarized light was 5\% with respect to that with the $p$-polarized light [Fig. 3(e)]. 
Here, a 100\% $s$-polarized ($p$-polarized) light excites only the anti-symmetric (symmetric) orbitals. 
Our laser system provides 97\% linearly polarized photons \cite{note_LP}, and thus the photoelectron intensity from the symmetric orbitals by the $s$-polarized light is about 2\% of the intensity by the $p$-polarized light in the experiment. 
It is also noted that the 7-eV photon in the present experiments mostly excites Cu 4$p$ orbitals, where the photoionization cross sections of 4$p$ orbitals are a few times larger than those of 3$d$ orbitals and are more than two orders of magnitude larger than that of a 4$s$ orbital \cite{Yeh_1985}. 
Consequently, one expects that the photoelectron intensity from the anti-symmetric orbital $d_{\rm yz}$ by the $s$-polarized light in the experiment is a few percent with respect to the intensity by the $p$-polarized light. 
The observed 5\% photoelectron intensity is roughly consistent with the sum of the above two contributions.  
Therefore, the non-reversal and the reduction of the spin polarization with switching the light polarizations are caused by the $p$-polarization component slightly included in our $s$-polarized light. 
One may observe the spin reversal with the perfect experimental geometry and a 100\% $s$-polarized light. 
Here, the optimization of the photon energy can enhance the photoelectron intensity from the $d_{yz}$ orbital that contributes to the spin reversal. 

The above discussion is based on the dipole approximation. 
Here, one may consider that the dipole approximation is not valid at these low energies \cite{Krasovski_PRB_2010} since the final states are not free-electron-like. 
The matrix elements of photoemission for individual light-electric-field components can strongly vary with the incident photon energy \cite{Kobayashi_PRB_2017, Bentmann_PRB_2017}. 
It is conceivable that the discrepancy of the photoelectron intensity and the spin polarization between the experiment and the calculations might be partly involved with the limitation of the dipole approximation. 

The total spin polarization of the surface states of Cu(111) should be nearly 1 since the anti-symmetric part is a little in the eigenstate. 
Here we note that the origin of the $L$-gap surface states of Ag(111) and Au(111) is quite similar to that of Cu(111); the symmetric (anti-symmetric) orbitals are dominant (minor). 
Therefore, the nature of the spin polarization and the spin-orbital entanglement of Ag(111) and Au(111) would be the same as that of Cu(111).

In summary, we have studied the $L$-gap surface states of Ag(111) and Cu(111) by laser-(S)ARPES. The spin splitting of the Ag(111) surface state is experimentally demonstrated for the first time.  
By using the three-dimensional spin detection, we have elucidated that the spin orientations of the Ag(111) and Cu(111) surface states agree with the conventional Rashba-type spin texture. 
In addition, we have experimentally revealed that the Rashba parameter of the Ag(111) surface state is 10$\%$ and 80$\%$ of those of Au(111) and Cu (111), respectively, being in good agreement with the theoretical predictions. 
Furthermore, we clarified that the symmetric orbitals are predominant in the surface-state wave functions. 

The authors thank Sogen Toyohisa for his support during the experiments. The authors also appreciate Ryo Noguchi for providing an Igor macro to draw the spin-resolved band images.  
The present work was financially supported by the JSPS Grant-in-Aid for Young Scientists (B) Grant No. 15K17675, and for Scientific Research (B) Grant No. 26287061.


\begin{references}
\bibitem{Rashba_para_calc} In the references 7, 8 and 23, the values of the $\alpha_{\rm R}$ are not described explicitly. Therefore, we calculate the $\alpha_{\rm R}$ with the formula of $\alpha_{\rm R} = \mathit{\Delta}\epsilon/2k_{\rm F}$, where $\mathit{\Delta}\epsilon$ is a size of the energy splitting and $k_{\rm F}$ is a Fermi wave vector.
\bibitem{note_LP} The degree of the linear polarization of the light is defined with the formula $P_{\rm LP} = (I_{x} - I_{y})/(I_{x} + I_{y})$, where $I_{x}$ represents one linearly polarized component and $I_{y}$ is another component perpendicular to $I_{x}$. Thus, the 97\% linearly polarized photons include the $I_{x}$ ($I_{y}$) component of 98.5\% (1.5\%). The linear polarization direction is continuously rotatable with respect to the light incident axis by using a half wave plate with conserving the degree of the linear polarization.
\end{references}
\end{document}